\documentclass[manuscript,screen, nonacm, table, acmsmall]{acmart}

\AtBeginDocument{%
  }

\acmYear{2025}
\acmDOI{XXXXXXX.XXXXXXX}

\acmConference[CSCW'25]{Proceedings of the ACM on Human-Computer Interaction}{October 18 — 22, 2025}{Bergen, Norway}
\acmISBN{978-1-4503-XXXX-X/18/06}
\usepackage{tabularx}
\usepackage{booktabs}
\usepackage{graphicx}
\usepackage{hyperref}

\usepackage{subcaption}
\usepackage{enumitem}

\usepackage{float} 
\usepackage{soul,color}
\usepackage[utf8]{inputenc}

\usepackage{longtable}
\usepackage[flushleft]{threeparttable}
\usepackage{chngcntr}
\usepackage{xcolor}
\definecolor{editCol}{rgb}{0.0, 0.0, 0.0}
\newcommand{\edit}[1]{{\textcolor{editCol}{#1}}}
\usepackage{pbox}
\usepackage{multirow}
\usepackage[colorinlistoftodos]{todonotes}
\usepackage{array}
\usepackage{makecell}
\usepackage{afterpage} 
\usepackage[title]{appendix}
\usepackage{subcaption,graphicx}
\usepackage{ragged2e}
\usepackage{comment}

\newcolumntype{L}[1]{>{\raggedright\let\newline\\\arraybackslash\hspace{0pt}}m{#1}}
\newcolumntype{C}[1]{>{\centering\let\newline\\\arraybackslash\hspace{0pt}}m{#1}}
\newcolumntype{R}[1]{>{\raggedleft\let\newline\\\arraybackslash\hspace{0pt}}m{#1}}

\begin{document}

\title[Youth Digital Privacy and Security Discussions Online]{Calculating Connection vs. Risk: Understanding How Youth Negotiate Digital Privacy and Security with  Peers Online}
\author{Mamtaj Akter}
\email{Mamtaj.Akter@nyit.edu}
\orcid{0000-0002-5692-9252}
\affiliation{%
  \institution{New York Institute of Technology}
  \city{New York}
  \state{New York}
  \postcode{10023}
  \country{USA}
}
\author{Jinkyung Katie Park}
\email{Jinkyup@clemson.edu}
\orcid{0000-0002-0804-832X}
\affiliation{%
  \institution{Clemson University}
  \city{Clemson}
  \state{South Carolina}
  \postcode{29634}
  \country{USA}
}
\author{Campbell Headrick}
\email{campbell.headrick@gmail.com}
\orcid{0009-0006-5704-3767}
\affiliation{%
  \institution{Vanderbilt University}
  \city{Nashville}
  \state{Tennessee}
  \postcode{37235}
  \country{USA}
}
\author{Xinru Page}
\email{xinru@cs.byu.edu}
\orcid{0000-0003-4067-7529}
\affiliation{%
  \institution{Brigham Young University}
  \city{Provo}
  \state{Utah}
  \postcode{84602}
  \country{USA}
}
\author{Pamela J. Wisniewski}
\email{Pamwis@stirlab.org}
\orcid{0000-0002-6223-1029}
\affiliation{%
  \institution{Socio-Technical Interaction Research Lab}
   \country{USA}
}

\renewcommand{\shortauthors}{Akter et al.}
\begin{abstract}
Youth, while tech-savvy and highly active on social media, are still vulnerable to online privacy and security risks. Therefore, it is critical to understand how they negotiate and manage social connections versus protecting themselves in online contexts. In this work, we conducted a thematic analysis of 1,318 private conversations on Instagram from 149 youth aged 13-21 to understand the digital privacy and security topics they discussed, if and how they engaged in risky privacy behaviors, and how they balanced the benefits and risks (i.e., privacy calculus) of making these decisions. Overall, youth were forthcoming when broaching a wide range of topics on digital privacy and security, ranging from password management and account access challenges to shared experiences of being victims of privacy risks. However, they also openly engaged in risky behaviors, such as sharing personal account information with peers and even perpetrating privacy and security risks against others. Nonetheless, we found many of these behaviors could be explained by the unique “privacy calculus” of youth, where they often prioritized social benefits over potential risks; for instance, youth often shared account credentials with peers to foster social connection and affirmation. As such, we provide a nuanced understanding of youth decision-making regarding digital security and privacy, highlighting both positive behaviors, tensions, and points of concern. 
We encourage future research to continue to challenge the potentially untrue narratives regarding youth and their digital privacy and security to unpack the nuance of their privacy calculus that may differ from that of adults.
\newline \textbf{Content Warning:} This paper includes example conversations of profanity and vulgar language for illustrative purposes. Reader discretion is advised.
\end{abstract}



\maketitle

\section{Introduction}
In today’s digital age, social media platforms, online gaming, and various other digital communication tools have transformed how youth interact and express themselves. Much of their social life now unfolds online, mediated by the internet \cite{boyd2008youth}. This digital landscape can be both empowering \cite{wong_hidden_2020} and challenging \cite{prievara_problematic_2019}. While some assume that immersion in online technologies \cite{gottfried_teens_2023, atske_about_2021} makes youth more tech-savvy than the adult population \cite{bennett_digital_2008}, scholars argue that they remain a particularly vulnerable demographic, frequently engaging in various risky behaviors~\cite{subramaniam2020examining, razi2022instagram}, such as viewing inappropriate content \cite{stevens_digital_2019}, encountering unwanted sexual solicitation \cite{drageset_social_2021, alluhidan2024teen, tanni_lgbtq_2024}, or sexting with strangers \cite{razi_Sext_2020, madigan_prevalence_2018}. Scholars often highlight a “privacy paradox,” in which youth overshare and engage in risky online behaviors despite potential privacy and safety risks \cite{barnes_privacy_2006, hargittai_what_2016}. Consequently, a more nuanced understanding of how youth engage online and manage their privacy is essential \cite{hodkinson2017bedrooms}.

Prior research has primarily focused on youth's individual awareness of privacy and security \cite{katavic2023digital} and examined various parental mediation strategies \cite{ghosh_circle_2020, hashish_involving_2014, akter_from_2022, akter_CO-oPS_2022} as well as educational programs \cite{chattopadhyay2019novel, zhang2017engaging, baciu2019adventures, katavic2023digital} aimed at increasing their awareness \cite{quayyum2021cybersecurity}. However, a significant body of networked privacy research \cite{zhang_privacy_2016, akter_evaluating_2023, akter2025examining, rashidi_Android_2018, Lin_expectation_2012, liu_understanding_2019} has demonstrated that knowledge and influence from social circles can play a crucial role in helping adult individuals improve their privacy and security awareness and behaviors - individuals frequently seek guidance from their close networks and learn from the informal stories shared within these circles \cite{redmiles2016think, mendel2017susceptibility, das_effect_2014, das_role_2015}.

In contrast, there is limited knowledge about how youth discuss privacy and security when interacting with their peers. Given that nearly half of U.S. teens report being constantly online, and seven in ten use social media platforms daily \cite{noauthor_teens_2024}, it is essential to understand how they exchange informal stories about privacy and security with their peers in these digital spaces. This body of existing research creates an interesting opportunity to explore whether youth themselves, as a tech-savvy generation \cite{works_you_2019, akter_it_2023, correa_brokering_2015}, engage in risky privacy and security behaviors and potentially are conduits of digital harm to others. Furthermore, to design support systems for youth, it is crucial to explore what motivates them to participate in such risks and share these incidents with peers.

Prior research \cite{razi_Sext_2020, hartikainen2021safe, prescott2017peer} demonstrates that analyzing digital trace data, such as private conversations, provides an opportunity to capture the unique dynamics of youth online interactions. To gain a deep understanding of youth privacy attitudes and behaviors, we examined private social media interactions among youth and their peers, focusing on conversations related to security and privacy. This approach offered a naturalistic way to observe youth perspectives, online behaviors, and challenges with privacy and security. 
Finally, we examined youth risk behavior related to privacy and security through the lens of privacy calculus  \cite{wisniewski_privacy_2022} in order to capture their unique process of risk-taking behavior. Privacy calculus is a prominent framework \edit{used to explain how individuals make privacy and security decisions by weighing the benefits against} the risks associated with their behaviors. \edit{In sum, our study sought to understand which perceived benefits outweigh which risks in regards to digital privacy and security, and how they navigate these issues in peer interactions. Specifically, we investigated how youth balanced maintaining social connections while protecting themselves in online spaces.} Our study was guided by the following research questions:

\begin{itemize}
\item \textbf{RQ1:} \textit{How do youth share and discuss their privacy and security experiences in private online chats?}  
    
\item \textbf{RQ2:} \textit{How do youth participate in risky behaviors regarding their digital privacy and security in private online chats?}

\item \textbf{RQ3:} \textit{How do youth weigh the risks against benefits of risky privacy and security behaviors?} 

\end{itemize}

\edit{Overall, our research questions focused on how youth discuss privacy risks, engage in risky behaviors, and balance risks and benefits in their decision-making.} To address these research questions, we conducted a study with youth participants aged 13 to 21. Using a custom web-based system, participants securely uploaded their private Instagram conversations.
This conversation dataset provided valuable insights into youth's personal online discussions on security and privacy topics. We used a set of privacy- and security-related keywords identified in prior research~\cite{wisniewski2018privacy, kumar2023understanding} to initially retrieve conversation data. Subsequently, we reviewed and sorted relevant discussions within this dataset. From the curated set of 1,318 conversations, we conducted a thematic analysis~\cite{braun2006using} to identify the themes.

Through the thematic analysis of private Instagram conversations, we identified key topics related to privacy and online risks, highlighting both protective strategies and risky behaviors. Our findings revealed that youth frequently discussed privacy threats, shared protective measures, and disclosed their password management practices (RQ1). However, many also engaged in risky behaviors, such as sharing login credentials with peers, and some even described perpetrating privacy violations, including hacking and scamming (RQ2). Our study further highlights the role of privacy calculus \cite{wisniewski_privacy_2022} in shaping youth decision-making (RQ3). While they recognized online security risks—such as poor password practices or financial scams—the social benefits of trust and peer approval often outweighed these concerns. For instance, youth knowingly shared passwords as a sign of trust, reinforcing social bonds \cite{marwick_networked_2014}. Additionally, some sought peer validation for engaging in privacy violations, illustrating how social influences drive both protective and risky behaviors.

Our study expands on the existing knowledge in CSCW by shifting the focus from individual digital privacy awareness to the peer-driven dynamics that shape youth behavior. Unlike previous research, which primarily examines educational programs or parental mediation, \edit{our analysis of private Instagram conversations provides a unique lens into how youth navigate privacy and security in real-world interactions. Our study offers novel insights into the specific topics youth discuss when addressing privacy and security concerns with peers, how they perceive and respond to online risks, and whether they engage in risky behaviors. Additionally, using the privacy calculus framework, we are able to understand how youth weigh privacy risks against social benefits in their digital interactions. Our findings have direct implications for the design of features such as controlled account-sharing options, enhanced privacy alerts, automated scam detection, and education-driven design elements that harness peer influence to foster safer digital habits.
} In summary, our study makes the following novel contribution to the CSCW research community: 

\begin{itemize}
    \item \edit{A comprehensive analysis of youth private conversations with their peers regarding their online security and privacy experiences.}
    \item \edit{An understanding of the social dynamics that shape youth's decision-making around digital privacy and security, including how they negotiate disclosures and engage in risky behaviors by themselves.} 
    \item Design recommendations for further steps to help mitigate youth participation in risky security and privacy practices, while allowing them to maintain social bonding with their peers.
\end{itemize}

\section{Background}

In this section, we introduce the relevant literature regarding peer influence in managing youth's experience with online privacy and security and challenging assumptions around youth privacy and security vulnerabilities, followed by privacy calculus as a theoretical lens to understand youth risky behavior. 

\subsection{Peer Influence on Youth Digital Privacy and Security Practices}
A large body of research has examined how individuals seek informal advice about digital privacy and security from their social circles. For instance, Rader et al. ~\cite{rader_stories_2012, rader_identifying_2015} found that people often acquire privacy strategies from the informal stories shared by family, friends, and colleagues. \edit{Exapanding on this, Das et al. demonstrated the critical role of social proof — awareness of how many friends use a particular security feature — in motivating individuals to adopt that feature} ~\cite{das_increasing_2014, das_role_2015}. In more recent work, Kropczynski et al. ~\cite{kropczynski2021examining} conducted a web-based survey to explore how tech caregiving within trusted circles, such as families, friends, and coworkers, enhances the collective ability of caregivees (those seeking support) to manage digital privacy and security.

\edit{Beyond informal advice-seeking, research has also explored how broader social processes influence privacy and security behaviors across different technology contexts.} For example, Emami-Naeini et al. examined the social factors influencing IoT privacy and security decision-making, discovering that participants were more likely to deny data collection by IoT devices when influenced by friends and made decisions more quickly when social cues were present ~\cite{emami_influence_2018}. Similarly, McDonald et al. explored how collaboration among loved ones can enhance individual's cybersecurity management ~\cite{mcdonald_building_2021}. \edit{In the context of mobile privacy and security, Wan et al. \cite{wan_appmod_2020} proposed a peer-based approach, titled AppMoD, allowing users to delegate their mobile privacy and security decisions to a trusted social connection who can be a family member or a close friend called "advisor". The trusted social connection can assist with the appropriate decision or make a decision by themselves on behalf of the user. Besides relying on the trusted individual, AppMoD also allowed users to review the other advisors’ prior decisions made for similar app permissions to help users make an informed decisions regarding their own mobile app permissions.}

While this above research has explored how social influence shapes adult individuals' privacy and security behaviors, much less attention has been given to peer support for youth in these areas. Studies indicate that, like adults, youth are also influenced by and learn from their peers. For instance, Lenhart et al. in ~\cite{lenhart2011teens} found that 18\% of U.S. youth considered their peers to be the biggest influence on appropriate internet behavior. Other studies have explored how youth seek peer advice on topics such as sexual health and mental health ~\cite{hartikainen2021safe, weinstein2016cope, alvarez2016online}. However, there is a significant gap in understanding how peer discussions and influence shape youth's digital security and privacy behaviors, including the types of digital privacy and security risks they discuss. Our research addresses this gap by investigating how youth discuss privacy and security with their peers on social media. Through a thematic analysis of private Instagram conversations, we examine how youth express their views on security and privacy and share their experiences with cyber threats.

\subsection{Challenging Assumptions around Youth Privacy and Security Vulnerabilities}
Most prior research on youth online safety has focused on how vulnerable youth are to digital privacy and security risks, often advocating for increased parental mediation, educational programs, or technological interventions to raise awareness and promote privacy-protective behaviors. Ironically, despite laws designed to safeguard children, such as the Child Online Privacy Protection Act (COPPA) \cite{noauthor_childrens_2013}, several studies have confirmed that many third-party mobile apps violate these regulations. For instance, Reardon et al. \cite{reardon_50_2019} \edit{analyzed over 88,000 mobile applications across different categories scraped from the U.S. Google Play Store and} found that numerous apps covertly access system resources (e.g., cameras, GPS) and collect personal information (e.g., contact lists, text messages, emails) without users' informed consent. Meanwhile, Akter et al. \cite{akter_from_2022} \edit{conducted a lab-based user study with 19 parent-teen dyads to explore how do they manage their mobile online safety, privacy and security within their families and found} that teens often install apps or grant permissions with little consideration for mobile privacy and security risks.

Given this, a significant body of research has explored strategies to improve youth privacy and security practices through education and intervention. Much of this work emphasizes parental mediation and its role in fostering positive parent-child relationships to promote safe online behaviors \cite{steeves2008closing, wisniewski2018privacy, badillo2020towards, park2024resilience, shin2016adolescents, akter2024towards}. For example, Hashish et al. \cite{hashish_involving_2014} introduced \edit{ an education-based mechanism titled,} "We-Choose," an app that encouraged collaboration between parents and children in selecting appropriate apps, leading to greater child engagement in their own privacy and security management. \edit{Their exploratory qualitative study revealed that participants felt it facilitated discussions and made the education more enjoyable and approachable for the children. Inspired by such open discussion-based learning,} Ghosh et al. in \cite{ghosh_circle_2020} developed "Circle of Trust," a mobile app allowing parents to guide their teens in identifying trusted contacts by reviewing risky content flagged by the app in text messages. 

In addition to parental mediation, several educational tools and programs have been proposed to promote youth online security and privacy \cite{chattopadhyay2019novel, zhang2017engaging, baciu2019adventures}. For example, workshops have been used to teach middle school students about visual digital privacy \cite{chattopadhyay2018middle}. 
There are also some programs designed for educating K-12 youth \edit{(children and adolescents enrolled in kindergarten through 12th grade)}, yet they are generally overshadowed by programs aimed at university students \cite{chattopadhyay2019novel}. 
While these programs and parental mediation strategies aim to protect youth from online privacy and security threats by enhancing their awareness, they often lack direct input from youth themselves \cite{kumar2023understanding}. In this study, we aim to explore this aspect by investigating what privacy and security matters youth discuss with their peers, whether youth exhibit any risk-seeking behavior, and what risky activities themselves they participate in themselves.

\subsection{Youth Risky Online Behaviors Through the Lens of Privacy Calculus}
\label{section3}
Instead of using the lens of youth being risk-seeking, we aim to understand their potentially risky behaviors through the lens of privacy calculus; if they are making potentially risky decisions online, we seek to understand why, based on the perceived benefits and risks. Thus, we approach our thematic analysis from the perspective of understanding individuals' online behaviors as a “Privacy Calculus.” In doing so, we use theories of social support to identify the benefits that factor into this calculus. Below, we describe these theoretical frameworks. 

\textit{Privacy Calculus.} A prominent framework for understanding people's decision-making process in regards to privacy is “Privacy Calculus \cite{wisniewski_privacy_2022}.” Through this lens, people's actions are shaped by weighing the benefits and the risks of performing the action. While people may express privacy concerns, often other factors outweigh these concerns. These factors include benefits from sharing personal aspects of their lives (e.g., achievements, emotions, and romantic relationships \edit{\cite{anderson2022post, bryce_role_2014}}), accessing advice for sensitive issues, creating meaningful social interactions \cite{rayland2023social}, and receiving emotional support by connecting with like-minded individuals without fear of stigma or judgment \cite{sanger2022social, vogels_teens_2023}. Privacy Calculus has been used to explain many seemingly paradoxical online behaviors \edit{where people engage in risky behaviors despite being concerned} \cite{gerber_explaining_2018}. By applying this framework to our data, we are able to gain insight into youth behaviors and attitudes towards privacy and security. Specifically, we go beyond identifying privacy and security risks and consider the benefits that youth perceive from engaging in those behaviors. Overwhelmingly, the benefits center on social support and so we also draw from theories of social support in our analysis.

\textit{Social Support.} Theories of social support focus on the extent to which actual and perceived support is given. A widely used framework is based on the work of House which distinguishes between four types of social support: Instrumental, Informational, Emotional, and Appraisal \cite{house_work_1981}. Instrumental support has to do with access to goods and services. This allows individuals to gain access to tangible resources and help from others. Information support consists of sharing useful knowledge with others. Learning about new and relevant information is a form of information support. Emotional support has to do with psychological constructs such as showing love, caring, or trust. Expressing empathy for what someone is going through is a common form of emotional support. Finally, Appraisal support occurs when others help the individual gain a more accurate appraisal of their current self and situation. For example, this can be in the form of affirmation or being reminded of their character traits \cite{house_work_1981}. This social support framework has been used to study a wide range of topics including impact on mental health \cite{laboy_differential_2014}, physical health \cite{noauthor_social_nodate}, and overall quality of life \cite{drageset_social_2021}. We use these types of social support to capture the range of benefits experienced by youth when engaging in or disclosing their privacy and security behaviors. Our results show that these types of support were driving factors of various risky privacy or security behaviors.  

\section{Methods}
\subsection{Study Overview}
\edit{We conducted a user study with U.S.-based youth aged 13 to 21, using a web-based survey.  Participants were recruited from across the United States through social media advertisements on Facebook and Instagram, as well as outreach to youth-serving organizations.} Participants first completed an eligibility screening questionnaire. To qualify, they needed to have maintained an Instagram account for at least three months and engaged in direct message conversations with at least 15 different people on the platform. Eligible participants were then presented with a consent form, and for minors, consent was obtained from a parent or legal guardian. Next, participants were asked to securely upload their Instagram private conversations through a custom web-based system we developed. \edit{We selected Instagram as our data source due to its widespread use among U.S. teens \cite{anderson2018teens}.} After submitting their Instagram data, participants completed a survey section to provide demographic information. A total of 195 verified participants contributed to the study, resulting in 32,055 Instagram private conversations. As compensation, each participant received a \$50 Amazon gift card for sharing their Instagram data and participating in the study. Our research was approved by the Institutional Review Board of the universities overseeing the study.
\subsection{Data Scoping and Relevancy Coding}

Our data scoping and relevancy coding process consisted of several steps. First, we selected keywords to conduct targeted searches for relevant messages. These keywords were drawn from previous studies \cite{wisniewski2018privacy, kumar2023understanding} that explored common terms used by youth when discussing digital privacy and security threats. For example, Wisniewski et al. found that terms like "hacked" and "stolen" frequently appeared in adolescent interviews about negative online experiences \cite{wisniewski2018privacy}. Similarly, Kumar et al. identified frequent mentions of "scam," "password," "security," and "privacy" in teen focus groups about common safety practices \cite{kumar2023understanding}. 

\edit{Building on these previous studies, we selected the keywords "hacked," "stolen," "scam," "password," "security," and "privacy" to guide our data scoping. Our initial search using these terms retrieved 2,591 conversations from \textit{n} = 149 unique participants.} Each conversation included the entire message history exchanged between the youth participant and others, spanning from minutes to years and covering multiple topics. To analyze these conversations, we divided them into sub-conversations centered around the keywords. For each keyword, we extracted the message containing the term along with the ten preceding and following messages to provide the necessary context for understanding its use. \edit{This process resulted in a total of 5,177 sub-conversations.}

The third author then reviewed the sub-conversations for relevance. Any sub-conversations containing the keyword but unrelated to technology or online contexts were marked as irrelevant. For example, a message like "The security guards will let people through, but the old lady at the front desk? Nah" was coded as irrelevant, as it lacked any connection to digital security and privacy, despite using the word "security." Other examples of irrelevant contexts included using "security" to refer to security personnel or \edit{"password" for sharing a temporary passcode for Zoom meetings.} After this relevancy coding, 1,318 sub-conversations remained, distributed across 741 conversations from 149 youth participants. Of these sub-conversations, 52\% contained the word "password," 23\% included "hacked," 12\% featured "scam," 7\% referenced "security," 6\% mentioned "privacy," and 1\% used the term "stolen." The total percentage exceeds 100\% because some sub-conversations contained more than one keyword.
\subsection{Data Analysis Approaches}
We conducted a qualitative thematic analysis ~\cite{braun2006using} to identify key themes in youths' discussions and behaviors around digital privacy and security. The third author, already familiar with the dataset from initial relevancy coding, began by coding a sample of 500 randomly selected sub-conversations to generate preliminary codes, with guidance from the first and last authors. Together, the first and third authors grouped these codes into cohesive themes organized by their relevance to the research questions. For RQ1, we identified themes that captured how youth discuss their experiences and perspectives on digital privacy and security issues. For RQ2, we focused on identifying themes illustrating potentially risky online behaviors. 
\edit{Next, the third author systematically coded the entire dataset, ensuring consistency in applying the initial codes. When new codes emerged, the researchers collaboratively determined their thematic placement. Upon agreement, the third author retroactively recoded prior sub-conversations to maintain consistency. This was an iterative process, where they constantly checked in with the other authors and formed a consensus.} 

\edit{For RQ3, we then conducted a deductive, top-down qualitative analysis ~\cite{bingham2023data} on the above dataset, applying the privacy calculus \cite{wisniewski_privacy_2022} and social support \cite{house_work_1981} frameworks, as described in \hyperref[section3]{Section 3}.} Specifically, we coded instances where youth and their peers discussed privacy and security risk experiences or participation in risky behaviors, noting 1) whether they exchanged social support in the form of a) love, care, and empathy, b) validation and affirmation, c) tangible resources, or d) useful information, and 2) whether the perceived social support outweighed the associated risks. \edit{This coding, led by the fourth author, helped us determine the extent to which social support benefits outweighed privacy and security risks for these youth.} 
\subsection{Participant Demographics}
\edit{In this study, we recruited 195 participants, of whom 149 engaged in privacy- and security-related discussions. These participants (\textit{n} = 149) ranged in age from 13 to 21, with a mean age of 17.26.} The majority (69\%) identified as female, with the remainder identifying as male (22\%) or non-binary/self-identified (9\%). Participant race was distributed as follows: 50\% identified as White/Caucasian, 26\% as Black/African-American, 22\% as Asian or Pacific Islander, 23\% as Hispanic/Latino, 4\% as American Indian/Alaska Native, and 3\% preferred to self-identify. (These percentages exceed 100\% as some participants identified as mixed-race.) Regarding sexual orientation, 52\% identified as heterosexual or straight, 26\% as bisexual, 11\% as homosexual or gay, and 11\% preferred to self-identify.

\subsection{Ethical Considerations}
We implemented comprehensive measures to ensure the ethical handling of sensitive data in our study. With Institutional Review Board (IRB) approval and all researchers completing both IRB and CITI Protection of Minors (POM) training, we prioritized participant confidentiality. Sub-conversations were documented without original usernames, allowing us to link messages to specific accounts if needed while maintaining anonymity. Within the codebook, usernames were replaced with generic labels to distinguish speakers without revealing personal information, and identifying details in messages were removed from all documentation outside secure storage. These details were substituted with generic terms, such as [NAME] or [PASSWORD], to maintain context. No images were included in the dataset, and insecure information like hyperlinks was replaced with placeholders. All data was stored on university-approved secure storage, with team members prohibited from saving it to personal devices or non-secure cloud platforms. Database and secure storage access was restricted to the university’s Virtual Private Network (VPN) for additional security.
\section{Results}
In this section, we present the findings on how youth discuss in online chats their their privacy and security experiences and participation in risky behaviors. We also present the themes related to how they weigh the benefits against these risky privacy and security behaviors.  
In the illustrative conversations presented in the sections below, youth participants are indicated by a ``P" while their peers are labeled with an ``O" or  ``O1/O2" in the case of multiple other participants. The demographic information provided refers to that of youth participants.

\subsection{The Nature of Privacy and Security Discussions Youth had on Private Online Spaces}
In this section, we discuss how youth shared and discussed their privacy and security experiences with their peers in Instagram private messages (RQ1). The most salient theme was when youth shared with one another about the potential threats they recognized and the ways they thwart these threats. Other themes where when youth discussed account login and password related matters, particularly cases where they themselves were victims of privacy and security risks.

\subsubsection{Youth discussed potential privacy and security threats they identified, and the protective measures they took to safeguard against the threats. 
} 

In the majority of conversations, youth discussed various potential security and privacy threats, such as scams and hacked accounts. They also frequently addressed other possible risks and personal privacy boundaries, demonstrating awareness of online privacy and security concerns. Notably, in many of these conversations, youth and their peers exhibited a keen ability to identify potential online privacy and security risks.  However, we also found instances of "false alarms," where youth or their peers initially perceived something as a threat to their security and privacy, only to later realize they were mistaken. In some cases, these misconceptions arose from assumptions about their peers that ultimately proved incorrect. In approximately 6\% of the conversations, youth specifically \textbf{discussed the hacked accounts they identified} (n=81). In most of these instances, they confidently indicated that they could discern when an account had been compromised, successfully avoiding the impersonators managing the hacked accounts. Typically, these hacked accounts belonged to someone the youth knew personally or followed on social media. The most common signs of a compromised account were receiving spam messages or unexpected advertisements. For example, youth frequently pointed out when their friends' Instagram accounts had been hacked, particularly when they began receiving suspicious or scam-related messages. Additionally, they identified compromised accounts by recognizing questionable links shared in message threads from friends or others. Interestingly, after noticing such suspicious messages, youth often initiated discussions with peers to confirm their suspicions about the security of a given account. Below is an example conversation where one peer sought confirmation from the group about a potentially dangerous link, which was quickly recognized and verified as a phishing attempt by the youth participant:
        \begin{quote}
            \textit{\textbf{O1 :} Sorry to bother y’all but do you know what this link is before I click it
            \newline \textbf{O1 :} [DELETED MESSAGE]
            \newline \textbf{O2 :} Was it hacked ?
            \newline \textbf{P :} it's a fake thing that takes ur password, don't click it
            \newline \textbf{O1 :} That’s what I though but wanted to clarify
        }
        \newline (Female, 16-year-old youth)
        \end{quote}


        Similarly, in approximately 5\% of conversations (\textit{n=69}), youth discussed encounters with \textbf{scammer accounts}, which they were able to identify in time, avoiding falling victim to these scams. In these discussions, youth shared experiences where malicious actors attempted to cause financial harm to themselves or others. They often grew suspicious due to sudden, unexpected "winnings" or unusual advertisements. In some cases, we observed that youth directly received harmful messages from third-party sources and immediately responded by identifying them as scams. In other instances, youth recognized a scam in group conversations and warned everyone about them. Yet, the majority of these conversations involved discussions with peers about whether a third-party account or its messages were, in fact, scams, as the following conversation illustrates: 
\begin{quote}
        \textit{\textbf{P :} I’m trying to tell whether it’s a scam or not but idk 
\newline \textbf{P :} Probably is but who knows
\newline \textbf{O1 :} whats her @
\newline \textbf{O2 :} she seems scammy
\newline \textbf{P :} Yeah ik and she just pmed what bank I have, if she asks for my routing number I’ll shut it down
}   
\newline (Female, 20-year-old youth)
    \end{quote}

Youth frequently discussed potential \textbf{digital privacy violations} (6\%, n=81) that they might expose themselves to through their online activities. These conversations often revolved around how their personal data is used, such as government surveillance and data harvesting. Some discussions reflected a general awareness of risks like information leakage and misuse, demonstrating the youth's understanding of digital privacy concerns. However, a predominant theme in these exchanges was the shared concerns about big tech companies stealing and misusing their information. For example, youth often expressed concerns that social media platforms not only have their data harvested by governments, but also engage in other forms of tracking. The following conversation illustrates such concerns, highlighting how youth critically reflect on digital privacy issues, demonstrating their skepticism about government surveillance and tech companies' data practices.
        \begin{quote}
        \textit{\textbf{P :} Why use tik tok and have your info be stolen and sent to China when you can use this app and have it stolen and sold to China (smile emote)
        \newline \textbf{O :} she acting like the US doesn’t track our every move
        \newline \textbf{P :} like we’re literally assigned a number at birth and ur scared of china seeing that a random person in florida is watching a dancing video on tiktok
        }
        \newline (Female, 18-year-old youth)
        \end{quote}

In some conversations, youth and their peers frequently discussed \textbf{personal privacy threats} (4\%, n=48) in broader terms. They shared their views on the extent of privacy violations they were comfortable with when interacting online. These discussions often touched on both the personal boundaries of individuals as well as broader considerations of what personal information is appropriate to share and with whom. In many cases, youth expressed a clear reluctance to share sensitive personal information, especially when interacting with peers or strangers online. The following example illustrates such an interaction, where a participant avoids disclosing personal details for privacy and safety reasons, highlighting how youth actively manage their privacy by withholding sensitive information when they feel it might compromise their safety online: 


        \begin{quote}
        \textit{\textbf{O :} Hold old are you turning? 
\newline \textbf{P :} I prefer to keep my age a secret, privacy and safety reasons. 
\newline \textbf{O :} ooOo okay
}  
\newline (Female, 19-year-old youth)
        \end{quote}
        
\subsubsection{Youth participated in discussions about their password management practices and challenges related to account access.}

In approximately 6\% of conversations, youth engaged in discussions regarding \textbf{their password practices} (n=80). In these instances, they showcased a variety of password strategies, including considerations of length, complexity, and the use of diverse characters. Youth also frequently talked about the password management applications they employed, how often they changed their passwords for particular accounts, and the importance of using different passwords for various accounts. Youth often demonstrated their understanding of strong password principles and how actively they applied them. Most of these conversations highlighted effective password practices, with peers boasting about the length and presumed complexity of their passwords, as the following example conversation illustrates: 

        \begin{quote}
            \textit{\textbf{P :} My password is 30 characters long lol
        \newline \textbf{O :} oh my god
        \newline \textbf{P :} 3 capitals
        }
        \newline (Male, 14-year-old youth)
        \end{quote}

While discussions often centered on overall password selection practices, another noticeable trend emerged \textbf{regarding forgotten passwords} (5\%, n = 66), which frequently caused issues with accessing accounts across various websites. In some conversations, youth and their peers explored methods for recovering accounts. However, in other instances, youth and their peers mocked each other for their forgetfulness or acknowledged having made similar mistakes. The following example illustrates this trend, where one youth admits to forgetting their password, and their peer playfully chastises them for not having better password saving practices:
        \begin{quote}
            \textit{\textbf{P :} I just forgot the password lmao
        \newline \textbf{O :} nice job
        \newline \textbf{P :} Ikr
        }
        \newline (Female, 18-year-old youth)
        \end{quote}


While discussions about passwords were the most common topic among youth, approximately 4\% of conversations focused on other \textbf{account access issues due to technical problems} (\textit{n}=56). In these instances, youth mostly faced technical problems, such as web server issues that locked them out of their accounts entirely. In some conversations, they discussed their struggles with the semantics of properly setting up their accounts for full functionality. In some of these discussions, youth and their peers discussed ways to navigate these issues. The following conversation exemplifies this pattern where the youth participants expressed frustration over not being able to access their account, another peer attempted to identify the problem. 
        \begin{quote}
            \textit{\textbf{P :} I have an account but it won't let me in. Tried to reset password but got no fucking email anywhere from the dozen times I've done it throughout the say so I had to create an account with my valencia email
            \newline \textbf{O :} You had to, its required by Valencia for each math class
            \newline \textbf{O :} Except for the one you did in person cause duh
            \newline \textbf{P :} I used the service, but I never had to do all this certification stuff
            \newline \textbf{O :} Well what did you use the service for then?
            \newline \textbf{P :} I don't remember
            \newline \textbf{O :} Hmm
            }
            \newline (Female, 18-year-old youth)
        \end{quote}

 Finally, in approximately 4\% of the conversations (\textit{n}=46), youth specifically discussed \textbf{different password practices they employed for shared accounts}. These conversations revolved around shared group accounts among peers, but often included discussions regarding what they considered appropriate security measures for accounts shared with others, covering topics such as which types of information or credentials were suitable to share with all account holders. In some of these conversations, we noticed that in an attempt to share personal accounts with each other, youth and their peers often exchanged their actual login credentials. The following exchange illustrates the youth and their peers' discussions regarding the complexity of a current shared password:

        \begin{quote}
            \textit{ \textbf{P :} can we change it to 15 so it’s much more easier haha
            \newline \textbf{O :} and maybe not all caps?
            \newline \textbf{P :} it’s not in caps
            \newline \textbf{O :} [PASSWORD]
            \newline \textbf{P :} i got in using lowercase
            \newline \textbf{O :} huh ok
            \newline \textbf{P :} can’t believe that this is our password (skull emote)
            }
            \newline (Female, 18-year-old youth)
        \end{quote}

\subsubsection{Youth shared stories of themselves being victims of privacy and security risks. 
}
At times, youth became victims of malicious actions online, suffering various consequences, and they often discussed these experiences. These conversations included general descriptions of the incidents, recovery measures taken, often with warnings to peers and other privacy and security risk experiences beyond the internet. 
In approximately 10\% of the conversations, youth shared their \textbf{stories of being scammed or hacked} (\textit{n}=123) with their peers, describing how their security and privacy had been compromised online. In these discussions, youth conveyed their experiences with a range of emotions, from alarm and agitation to lighthearted humor. Many conversations began with the victim explaining what had happened, although in some cases, the topic arose after a peer mentioned suspicions of a scam or hack. An example of such response to a security breach is illustrated below, where a participant shared the incident with a peer after finding out that their Spotify account was hacked:

        \begin{quote}
            \textit{\textbf{P :} I THINK THEY HACKED MY SPOTIFY
            \newline \textbf{P :} I WAS SCARED
            \newline \textbf{O :} thank god i do zaful i-
            \newline \textbf{O :} WHAT
            \newline \textbf{P :} BUT IDK IF IT WAS that
            }
            \newline (Female, 18-year-old youth)
        \end{quote}

Among the conversations where youth shared their experiences of being scammed or hacked, youth often specifically discussed \textbf{the recovery and protective measures they took following these incidents} (6\%, n=78). Youth occasionally employed various strategies, such as reporting the hacked account or the individuals responsible for compromising their account. In some cases, they also took the initiative to prevent similar incidents from happening to their peers, offering warnings about their hacked accounts. This often involved apologizing and advising others to ignore any suspicious messages sent from their compromised accounts. However, the most common recovery method youth used was changing their passwords or creating new accounts, as the following example conversation illustrates:
        \begin{quote}
            \textit{ \textbf{P :} I did not send that
            \newline \textbf{O :} Thats why I'm confused
            \newline \textbf{O :} Lol
            \newline \textbf{P :} Looks like it’s time to change my password again
            \newline \textbf{O :} Yes sir
            \newline \textbf{O :} Someone hacked ur shit
            }
            \newline (Male, 20-year-old youth)
        \end{quote}

Overall, we found that youth discussed their privacy and security practices and experiences with their peers. This included messaging about the risks of potential threats, their individual security and privacy practices, as well as discussing the incidents where they had been victims of security and/or privacy violations. In many of these cases, youth and their peers would discuss ways to resolve issues that had occurred. 

\subsection{Youth's Privacy and Security Behaviors in Private Online Spaces}
    In this section, we explore the themes related to how they engaged in risky behaviors in conversations where they discussed privacy and security matters with their peers. Below, we delve into these themes, such as youth's sharing accounts and login credentials, as well as more serious actions like hacking and scamming others in greater detail.

\subsubsection{Youth shared sensitive account information with others 
}

In approximately one-fourth of the total conversations, youth took part in risky behavior with others. These conversations included sharing account passwords for temporary purposes and discussions about sharing their accounts with peers and family members. Specifically, in around 16\% of the conversations, youth directly \textbf{shared account passwords} with peers (\textit{n}=215). In these conversations, youth temporarily shared account access with peers, often in one-on-one conversations. However, in some instances, we noticed that the passwords and sensitive account details were shared in group chats. In the following example conversation, we see such patterns where youth participant posted their account credentials with multiple people. 

        \begin{quote}
            \textit{\textbf{O1 :} Hey y’all even tho I said I can’t admin anymore, I saw BNHA on the recommendations and I’d love to post about BNHA whenever! If you guys need the extra hands.
        \newline \textbf{O1 :} My phone isn’t letting me log out of accs so I’m still logged in
        \newline \textbf{O2 :} OoF fam we just got three new admins we’re working on to get the password. I’ll have to see?
        \newline \textbf{P :} If you need it I’m pretty sure the password is [PASSWORD]
        }
        \newline (Female, 19-year-old youth)
        \end{quote}

           While the above conversations focused on specifically disclosing passwords with others, there were instances where youth generally discussed \textbf{sharing their accounts with others} (7\%, n=90). Youth often shared accounts with their peers to collaborate on shared ideas or projects, enhance communication and work processes, manage tasks and ensure everyone has access to the necessary files and information. Therefore, the majority of these conversations took place in group chats where youth primarily shared accounts with their friends and others. A few conversations that were one to one had discussions of youth's sharing accounts with their family members. This often occurred with parents to allow for monitoring, or merely shared access to services or paid accounts between family members. 
           

           
        \begin{quote}
            \textit{ \textbf{O1 :} ok im dragging 3 more ppl in
        \newline \textbf{O2 :} alright bros so we somewhat got people coming lol
       \newline \textbf{P :} The password worked for a while, the previous owner of the account reset the password
       \newline \textbf{P :} I tried that password and it didnt work
       \newline \textbf{O2 :} who was the previous owner?
       \newline \textbf{P :} [NAME]
       \newline \textbf{O2 :} oh lol then we should text her
       }
       \newline (Female, 19-year-old youth)
       \end{quote}

Around 5\% of conversations revealed instances where youth expressed \textbf{intentions to share passwords in the future} (\textit{n}=69). These cases typically involved youth offering to share account passwords later or requesting it from others, rather than immediate sharing. Interestingly, such intentions were more frequently observed in group chats rather than in one-on-one conversations. The following example features a group of peers who had agreed to share an account as admins, but had not yet exchanged the necessary information due to one person forgetting the password. This type of discussion was common, with one peer requesting the account details and the youth participant offering to provide it later:
        \begin{quote}
            \textit{\textbf{O :} I also need information for the account because I can’t really post without a passwordjdjddjjdjssjsj or is that not how we’re doing this?
        \newline \textbf{P :} Nah you’re gonna be an admin
        \newline \textbf{P :} I forgot the password anyways
        \newline \textbf{P :} Sorry
        \newline \textbf{O :} Damn
        \newline \textbf{P :} Someone else should know it??
        \newline \textbf{P :} I hope
            }
        \newline (Female, 19-year-old youth)
        \end{quote}

\subsubsection{Youth perpetrated security and privacy threats to others (10\%, n=135)}
Another form of risky behavior exhibited by youth involved actively compromising the security and privacy of others. These threats included both past actions and future plans to target individuals and businesses, typically without facing any significant consequences. In approximately 8\% of the conversations, youth engaged in malicious behavior, often discussing attempts to \textbf{hack or scam individuals online} (n=105). In most cases, the targets of these attempts were specific individuals, typically people the youth knew personally. Occasionally, youth also sought to access product-related information from local businesses. There were a few instances where the acquisition of pirated software was referred to as "hacking." While the majority of these conversations involved discussing actions that had already taken place, some revealed intentions to carry out such activities in the future. \edit{The following example illustrates a case where a youth admits to hacking the account of someone they knew of:
}
        
\begin{quote}
\edit{    \textit{\textbf{P :} got into his account last night.  
    \newline \textbf{O :} No way, how?  
    \newline \textbf{P :} dude had the most basic password .  
    \newline \textbf{O :} LMAO, what’d you do?  
    \newline \textbf{P :} nothing wild, just peeked around and hit up his friends.  
    \newline \textbf{O :} LOL, what’d you say?  
    \newline \textbf{P :} just enough to make him look goofy.}  
    \newline (Male, 16-year-old youth)}
\end{quote}

        We also observed that participants sometimes \textbf{engaged in spamming and/or scamming} in group chats (2\%, n=30). In these conversations, youth posted various malicious messages disguised in product promotion, service advertisement, or  personal account, often including attached link. Youth also sometimes sent these messages repeatedly in quick succession. Notably, other group members rarely responded or reacted to these messages, suggesting that youth could engage in such behavior with little resistance or consequence. 
        The conversation below provides an example of youth spamming in group chats:
        \begin{quote}
            \textit{\textbf{P :} [LINK]
        \newline \textbf{P :} All u have to do is click the link and it will \newline give me \$10 so pls do! This is not a scam thing either
        \newline \textbf{P :} Subscribe to her channel! Just made a new video btw! [LINK]
        \newline \textbf{P :} Subscribe to her channel! My bestie just made a new video and I'm tryna promote it for her! [LINK]
        \newline \textbf{P :} [LINK]
        \newline \textbf{O1 :} Hello (wave emote) I’m so excited right now
        \newline \textbf{O2 :} Hello
            }
            \newline (Female, 18-year-old youth)
        \end{quote}

As such, youth not only discussed their privacy and security experiences but also engage in risky behaviors through private channels on social media. They sometimes exchanged sensitive information on shared accounts. Other times, they engaged in high-risk behaviors such as hacking or scamming others online. Below, we unpack how youth weighed benefits and risks associated with these unsafe privacy and security practices.

\subsection{The Ways Youth Weighed Risks Against Benefits of the Risky Privacy and Security Behaviors}
The previous section illustrates how youth sometimes engaged in unsafe privacy and security practices. In this section, we explore how their behavior was often driven by a privacy calculus where they balanced various types of social support against using and disclosing the above unsafe privacy and security practices. We observed the following types of privacy calculus.

\subsubsection{Youth disclosing privacy and security mistakes to seek \edit{emotional support while offering and receiving informational} support.} 

In many conversations, we observed that youth often admitted their own privacy or security mistakes to peers. \edit{Sometimes they would receive instrumental support in the form of ideas and advice from others. For example, this youth turned to his friend for guidance and support after clicking on a phishing link: 
 \begin{quote}
            \textit{\textbf{P :} Not again, Im locked out. I think I got hacked. 
\newline \textbf{O:} Yikes. Did you try resetting your password? You gotta stop clicking on sus stuff.
\newline \textbf{O:} Turn on that two-factor. 
\newline \textbf{P:} Its not letting me reset.
}
  \newline (Male, 17-year-old youth)
\end{quote}
}
Interestingly, in the majority of those instances, they \edit{were also trying to provide informational support by sharing} these experiences to make others aware of potential risks. 
In the example conversation below, the youth disclosed how a malicious actor gained control of their own account, potentially putting others at risk. The original poster shared this incident as a form of informational support to help others avoid the same security error. Although their peer couldn’t offer direct assistance to fix the situation, they provided reassurance, indicating it’s okay to make mistakes. 

       \begin{quote}
            \textit{\textbf{P :} I'm sorry I got hacked change you password right now if you logged into that website!
        \newline \textbf{O :} i didn’t lmao
        \newline \textbf{P :} I'm really sorry! I copy and pasted that message just in case someone one else was a big gullible idiot like me...
        \newline \textbf{O :} nooo it’s okay (Crying emote x2)
            }
            \newline (Female, 16-year-old youth)
        \end{quote}


Rather than judging, we observed that in most of these conversations, youth's peers tended to be understanding and empathetic, showing how sharing such experiences can be a valuable source of emotional support for youth. We often noticed that youth shared stories of their accounts being hacked or experiences of being scammed as a way to seek reassurance and empathy from their peers—a form of emotional support. Often, youth recounted how they fell for online financial traps, such as fraudulent shopping sites, and expressed their concerns with friends once they realized they may have been scammed. In response, peers typically offered reactions of unease or worry, showing support for their friends in distress.


\begin{quote}
\edit{\textit{\textbf{P :} Bruh, I think I got scammed… ordered something online, and now there’s no tracking number or anything :=
\newline \textbf{P :} feel so dumb for falling for it. Probably never seeing that money again.
\newline \textbf{O:}  Ugh, that’s the worst! Hope you can get a refund or something. Don’t stress too much, scams be getting everyone these days.}
\newline (Male, 18-year-old youth)}
\end{quote}

\subsubsection{Youth's questionable privacy and security practices \edit{as a way to} obtain instrumental support.} 


In our RQ2 results, we observed that youth often engaged in questionable privacy and security practices, such as password sharing. However, these practices were primarily a means to obtain practical, instrumental support from others. In many of these conversations, we noticed that password sharing served a functional purpose, such as enabling multiple individuals to administer a shared account. The example conversation below illustrates this pattern, where sharing an account password allowed youth to achieve the practical objective of distributing administrative access among multiple individuals, rather than limiting it to a single administrator.

\begin{quote}
\textit{P : Welcome new Instagram admins!! After a little more organization we’ll give you to password to access the account, and find out what days are best for you to post. There’s another, separate chat that consists of all the old/current admins. The other chat is mainly consisted of random content and isn’t all that active but feel free to ask to join in on the chaos whenever. Welcome!!
 \newline O1 : Welcome everyone (pink hearts emote) I’m [NAME]!! I can’t wait to work with you guys on the account
 \newline O2 : Thanks for choosing me as an admin (heart emote)
 }  \newline  (Female, 19-year-old youth)
\end{quote}


However, in some other conversations, we observed youth engaging in account password sharing that did not appear to be for joint administration purposes. In these cases, the reason for granting access was unclear; however, it was evident that passwords were being traded as part of a transactional exchange, a form of instrumental support, treated almost as commodities.


        \begin{quote}
            \textit{\textbf{P :} You need to tell me more things
        \newline \textbf{P :} [NAME] did that kinda exept shorter and I got some things wrong but then she kinda told me some answers and so I did it again and got it close enough and the deal was she'd give me her password so now I know her password
        \newline \textbf{O :} Nice job! Lol
            }
            \newline (Female, 16-year-old youth)
        \end{quote}



In some other instances of password sharing, youth did not intend to grant others complete control over their accounts but instead shared their personal credentials to obtain specific help. For example, in the conversation below, a peer shared their Apple ID information with a youth participant to help locate their lost phone. 

        \begin{quote}
            \textit{\textbf{O :} Can you u use find my iPhone for me
        \newline ...
        \newline \textbf{P :} yeah sure
        \newline \textbf{P :} dont i need you email and password tho
        \newline \textbf{P :} bc i don’t have your location on
        \newline \textbf{O :} YE I’ll give u my Apple ID [NAME]
        \newline \textbf{O :} [USERNAME]
        \newline \textbf{O :} [PASSWORD]
        }
        \newline (Female, 18-year-old youth)
        \end{quote}

\subsubsection{Youth shared sensitive information to reinforce \edit{trust}, \edit{a form of emotional support}.} 


In addition to engaging in risky privacy and security practices, we also observed that youth often shared sensitive personal information to build \edit{trust and stronger social connections with others, a form of emotional support}. 
Most frequently, this included sharing account credentials. In the example conversation below, not only did the youth share her own password, but she also revealed that she and the male friend she referenced use the same password. Additionally, she granted another close friend access to her phone by enabling biometric access through a fingerprint. The peer’s reaction to this disclosure was telling how they express feeling left out, indicating that withholding sensitive information can carry negative social implications within peer relationships.

\begin{quote}
\textit{P : And [NAME] knows my password to everything 
\newline O : I-
\newline P : and my phone password is the same as his so
\newline P : and [NAME] is special
\newline O : okaY more hurt
\newline P : so she gets her fingerprint
}  
\newline (Female, 16-year-old youth)
\end{quote}

Additionally, in many instances, we noticed how youth frequently disclosed common background details as a way to foster social bonds. As the example conversation below shows, the youth placed themselves in a vulnerable privacy position by revealing their alternate account name. Although they initially took steps to dissociate their identity from that account, they openly connected it to their true identity in this interaction. The individual also shared potentially sensitive information about their sexual orientation.

        \begin{quote}
            \textit{\textbf{P :} I’m [NAME], [PRONOUNS], also a lesbian! :))
        \newline \textbf{P :} I use [FALSE NAME] on that acc for privacy u know
        }
        \newline (Non-binary, 13-year-old youth)
        \end{quote}


\subsubsection{Youth bragging about bad privacy and security practices to gain \edit{appraisal support}.}

    
We saw many instances where youth did not just discuss unsafe privacy and security practices, such as frequently using the same password for a variety of different accounts, or having never changed passwords over an extended period of time, they bragged about them with their peers. As in the example conversation below, the youth's peer offered social affirmation, \edit{a type of appraisal support,} rather than improved password security advice or condemnation. There were many examples of sharing such insecure practices that were met by peer understanding and support. 
        
        \begin{quote}
            \textit{\textbf{O1 :} i change my pas every 2748383 years
            \newline \textbf{O2 :} I have a lot but yeah I do I just didn’t know which was which
            \newline \textbf{P :} i change my password never
            \newline \textbf{O2 :} LMAO
            }
            \newline (Female, 17-year old-youth)
        \end{quote}

Surprisingly, in some of these conversations, we noticed that peers continued to share strong social affirmation even when youth shared that they were the perpetrators of privacy and security threats towards others. As the example conversation below shows, the youth shared how they violated another person's privacy by preventing access to their own account. The strong social affirmation expressed by the peer legitimized this practice despite it being a security breach. These examples illustrated how the importance of safe and fair privacy and security practices varied greatly depending on purpose of such practices. In fact, security violations were outweighed by the strong social agenda to impede someone's ability to access their account. 

        \begin{quote}
            \textit{\textbf{P :} I purposely hacked him lol
            \newline \textbf{O :} LOLOLOLOLOLOL
            \newline \textbf{O :} THIS GENIUS HAHAH
            \newline \textbf{P :} And changed his password to something really different
            \newline \textbf{P :} So he’s locked out of his account…
            }
            \newline (Female, 17-year old-youth)
        \end{quote}

\section{Discussion}
We now discuss the extent to which youth were aware of privacy and security risks, and their practices based on this awareness. We also reflect on the social support benefits of disclosing their privacy and security experiences and behaviors to peers online. We conclude with implications for design and education. 

\subsection{Youth Privacy \& Security Awareness and Protective Actions}


A major trend we observed was that youth frequently recognized potential threats to their privacy and security, both before harm occurred and after breaches. They were also aware of issues related to digital surveillance by social entities and government bodies. These findings align with prior studies suggesting that youth take digital privacy seriously and are knowledgeable about various threats \cite{madden2013teens, lenhart2011teens, boyd2011social}. Additionally, Akter et al.~\cite{akter_from_2022} demonstrated that teens, being tech-savvy and cautious of malicious intent of the third party mobile apps, were particularly concerned about the app permissions granted to the apps on their mobile devices. In our study, we saw how youth often took specific protective measures to mitigate the risks they encountered. 
While societal assumptions and some previous research \cite{steeves2008closing, wisniewski2018privacy, shin2016adolescents} suggest that youth need parental mediation for online protection, our findings indicate that they already possess significant awareness and protective knowledge. We therefore recommend that future research focus on guiding youth toward advanced protective practices that leverage their existing knowledge, rather than solely designing technologies to protect them. Furthermore, rather than viewing youth strictly as a vulnerable group, researchers could explore how their knowledge and strengths \cite{park_towards_2023} might support peers or even family members, as suggested by many networked privacy research for adult populations \cite{wan_appmod_2020, akter_evaluating_2023}.


We also observed that youth demonstrated a similar ability to identify risks associated with sharing personal information. When others, especially strangers, asked personal questions on private chat, youth frequently declined to answer, indicating a strong sense of personal privacy and awareness of the risks involved in sharing private information. 
However, we also noted a contrasting behavior: youth frequently shared their personal account information, including passwords, with peers. This suggests that their sense of privacy may depend on the nature of their relationship with those requesting the information, \edit{which aligns with Nissenbaum's framework of contextual integrity \cite{nissenbaum_privacy_2004}. Contextual integrity is a privacy framework often used to analyze how individuals navigate privacy trade-offs based on situational expectations and norms, providing a valuable lens for studying digital privacy behaviors \cite{wisniewski_privacy_2022}.}
This finding also aligns with previous research \cite{akter_from_2022, akter_it_2023}, which showed that individuals are more willing to share personal information with trusted family and friends. 
\edit{The concerning trend we observed in our study was the youth's personal information-sharing practices as a mechanism for building trust even before establishing that the person should be trusted. This is in line with prior work which showed how young people assess risk and trust in online interactions, particularly concerning the sharing of personal information. 
Despite youth being well aware of the potential risks associated with their online activities, many viewed sharing personal details as essential for building online relationships~\cite{bryce_role_2014}. } \edit{Taken together, our findings suggest the need for support mechanisms for youth to help evaluate trustworthiness and verify identities in digital spaces.} 

\subsection{Privacy Paradox and Youth Peer Support}
Our findings reveal notable discrepancies between youth’s stated privacy perceptions and their actual behaviors, which may appear as a "privacy paradox" ~\cite{norberg2007privacy, phelan2016s} (e.g., recognizing the importance of secure passwords while still sharing them with peers). However, examining these behaviors through the lens of privacy calculus suggests underlying motivations. The known risks are often outweighed by perceived social benefits. For instance, sharing passwords, while risky, enables youth to build trust\edit{, a form of emotional support,} during a critical period of psychosocial development, where forming connections outside their family is a top priority. What might seem like an overly risky willingness to share personal information on social media ~\cite{barnes_privacy_2006} may actually reflect a calculated decision-making process in which social benefits are prioritized over potential risks. 
Thus, it is essential for researchers and designers to consider this broader context when evaluating youth privacy and security practices. Regardless of a youth’s understanding of risks, these social benefits may outweigh them. We urge researchers to take into account the powerful influence of peers and social pressures, which are especially impactful at this stage of development.

Another key finding was that youth play a unique role in providing \edit{both informational and emotional support} to one another when experiencing privacy violations. They can admit mistakes to peers without fear of harsh judgment or punishment—often a concern in relationships with parents or other adults, where oversight and authority are more prominent. Even when peers \edit{give informational support by pointing} out errors, such as forgetting a password and not storing it safely, the feedback is typically a light-hearted reminder rather than a reprimand. This peer-based support may be essential for youth, allowing them to reflect on mistakes and consider alternative actions. This aligns closely with the Reflective Learning method, widely used in education, where students review past errors to inform future decisions. Thus, sharing mistakes with peers and receiving constructive feedback may foster better privacy practices over time. Although our data did not track whether youth changed their behaviors as a result of this reflection, we encourage future research to explore the effectiveness of \edit{this approach of providing informational support in an emotionally supportive way}.

\subsection{Social Support in Youth Risk-Taking Behaviors}

While many youth demonstrated strong security and privacy practices, we also saw youth engaged in various risky online behaviors \edit{in order to gain access to one of the four types of social support (informational, instrumental, emotional, appraisal). Thus, we need to pay attention to all of these forms of social support in order to identify the key motivators behind youth privacy and security decisions. However, not all behaviors were equally risky. For instance, exchanging passwords and} sharing impersonal accounts or family accounts can offer practical benefits \edit{ and may not be inherently risky. } Nevertheless, sharing sensitive passwords with those outside of the the account owner(s) and leaving digital traces of such information poses significant risks. Privacy and security threats on social media continue to evolve with a recent report documenting over 26 billion data breaches involving youth on popular platforms like Twitter~\cite{breach2024}. Private online chats, such as Instagram direct messages, are equally vulnerable~\cite{Instagrambreach2020}. Further research is required to better understand the \edit{ extent of and the} risks associated with youth password and account sharing, as well as to develop interventions that raise awareness of these risks\edit{, while accounting for the instrumental support youth need. Teaching youth safer ways of accomplishing the same tasks could provide them with a less risky alternative.}

Additionally, a concerning trend was observed in which youth engaged in privacy and security threats against others for amusement. Prior research has noted similar patterns of youth involvement in such malicious actions~\cite{kim2022theoretical, back2019youth}. This behavior is particularly troubling given the apparent lack of remorse among youth for these actions~\cite{hu2012moral}. Educational initiatives are therefore needed to highlight the negative consequences of these behaviors on both themselves and others, encouraging more responsible conduct. Equally disconcerting is the minimal social repercussion from peers when youth disclose such actions. \edit{In fact, youth offered appraisal support in the form of affirming} these behaviors. Given that adolescents are especially sensitive to peer pressure and social approval, the normalization of these behaviors is problematic. While adolescents can assess risks, they often exhibit heightened sensitivity to peer validation in risky scenarios~\cite{smith2014peers}. Establishing strong social norms and clear guidelines that discourage risky privacy and security behaviors is critical in mitigating these actions among youth.

\subsection{Implications for Design and Education}
Our results confirmed both positive and concerning trends in youth's discussion of privacy and security matters through private online chat. In this section, we provide implications for design and education to promote online privacy and security of youth while maximizing benefits and minimizing the negatives of online peer interaction regarding privacy and security topics.

\subsubsection{Design Implications}
Based on our findings, we suggest the following design steps to provide outlets for stronger youth security and privacy practices:

\begin{itemize}
    \item \textbf {Implementation of a Moderated Forum for Youth Privacy and Security:} In order to fully leverage the strengths of peer-to-peer youth support, designers can consider implementing a dedicated help forum for discussing security and privacy practices. 
    Moderation mechanisms should be in place to make sure that shared information is accurate and discourage behaviors such as inflicting privacy violations on others.
    
    \item \textbf {Real-time Nudges as Safe Sharing Reminders}:  \edit{Prior work has shown that real-time support can serve as an effective tool for helping teens learn about safe behaviors~\cite{agha2024systematic}. Accordingly, there has been an increase in such real-time approaches, where social media platforms are testing nudges for safety~\cite{ghaffary_instagrams_2022}.} Therefore, when youth attempt to share sensitive information (e.g., passwords), a real-time prompt could remind them of safer practices and offer quick options for trusted sharing alternatives (e.g., “Share access temporarily without sharing your password”).

    \item  \textbf{Automated Spam Recognition and Warnings for Risky Content:} While the youth were able to recognize instances of suspicious messages, there were several times when messages would be spammed and not commented on. As many of the potential forms of hacking or scamming on social media involve tricking others into clicking on malicious hyperlinks, the Implementation of automated spam recognition to detect and provide warnings in conversations such as these could help alert the conversation participants. \edit{Recent advancements in Large Language Models have made automated spam recognition even more promising (e.g.,~\cite{chang2024site, halder2024enhancing, schesny2024enhancing}). } 
    
    \item \textbf{Account Sharing Features with Granular Security Settings:} While some account-sharing practices are benign, \edit{we acknowledge that such practices can potentially lead to serious privacy and security risks. } To support youth to achieve a level of trust building without completely sacrificing security in account sharing, designers can provide a partial account-sharing feature that allows multiple users to share access to content without having to share a password,\edit{ which has been successfully implemented in smart home devices~\cite{alghamdi2024share}.} With this feature, a user could grant access to another user to see content and use a subset of features, yet, for security reasons, that access could be revoked in the future and the other person cannot see or change the account owner's password; a safeguard could be allowing the account owner to “undo” any action performed by other users who were given access. \edit{A similar feature can be implemented on social media platforms.} 
    
\item \textbf{Incentives for Spreading Awareness of Risks:} One potential method of increasing awareness of security and privacy risks among youth would be providing some formal incentive attached to spreading resources. Social media platforms can consider incentivizing accounts through social features (e.g., visibility of accounts) for sharing specific security and privacy-related resources to motivate youth to share such resources.

\end{itemize}


\subsubsection{Implications for Education}
Beyond design implications, below we suggest opportunities for educative resources to further address forms of risky behavior youth engaged in:

\begin{itemize}
    \item \textbf {{Raising Awareness of Negative Consequences of Privacy and Security Risks among Youth}}: A recurring practice youth discussed is piracy of internet software, which can prove dangerous due to the untrustworthy nature of distributors of such software. While educational resources can be found online about the subject, they are relatively out of the way, and the topic could potentially receive greater attention as one to educate youth about beyond topics such as password practices and credential sharing.
    In a similar vein, while the focus is dedicated to protecting youth online, fewer resources discourage outright malicious behavior. Similar to the way topics such as cyberbullying are covered, youth could benefit from greater discouragement from involvement in security and privacy threats. 
    
    \item \textbf {Gamified Security Practices with Peer Recognition}: Educators can create gamified challenges that promote security practices, allowing users to earn badges for good behaviors (e.g., creating strong passwords, enabling two-factor authentication). Allow peers to see and react to these badges, reinforcing positive behavior. Given that youth in our study often engage in risky behaviors for peer affirmation, gamification, and social recognition can shift this drive toward safer practices by making security efforts visible and valued in their social circles.

    \item \textbf {Peer Community Norm-Building Programs}: We can also design educational programs that allow youth to create and define peer community norms for privacy and security, allowing them to set boundaries and expectations within peer groups. Youth often seek peer affirmation and may act in ways that compromise security for social gain. By co-creating norms, they can build a culture where safe practices are socially reinforced and risky behaviors are discouraged.

    \item \textbf {Integration of Educative Features on Social Media}: Finally, social media can help educate youth about privacy and security practices through tutorials or pop-up tips that users can access directly within social interactions. For example, when sharing credentials or personal information, the system could push tutorials that suggest safer alternatives or remind users of potential risks. Youth often share sensitive information with peers due to social trust. On-platform educational content that is triggered in the moment of potentially risky behaviors can encourage thoughtful choices, leveraging social moments as learning opportunities.

\end{itemize}

\subsection{Limitations and Future Research}

While our research demonstrated a nuanced understanding of youth security and privacy practices, there are some limitations worth acknowledging. 
The dataset consisted of private conversations donated by English-speaking US youth, thus only demonstrating the practices of a specific demographic, rather than all youth on Instagram. 
\edit{Our participants were required to be active Instagram users for a certain period and voluntarily donate their data for our study. As a result, our findings may not fully generalize to other youth populations. Future research could examine diverse perspectives by collecting conversation data from youth across different social media platforms. Additionally, while our study analyzed data from individuals aged 13 to 21, we recognize that this range spans from adolescence to emerging adulthood. Future work could investigate whether generational shifts occur over time, offering deeper insights into evolving privacy and security behaviors.}

Furthermore, the perspectives and experiences youth share on Instagram do not necessarily demonstrate the full scope of their experiences. Youth may choose not to share certain risky behaviors, for example, or otherwise not be interested in describing the full scope of their security and privacy experiences. Furthermore, sub-conversations do not always carry the full context of the interaction, as at times messages were deleted, or the conversation context attached to certain messages was ambiguous. Even with the limitations stated above, Instagram conversations did provide a unique look into the lives of youth and their security and privacy practices.

\section{Conclusion}
Our study sheds light on the complex, often contradictory ways youth navigate digital privacy and security, balancing social connectivity with risky behaviors. While youth demonstrated awareness and proactive strategies, the lure of social validation often led to sharing passwords, sensitive information, at times, even participate in security threats against others. This peer-driven dynamic underscored a “privacy calculus,” where social benefits often outweighed their perceived privacy risks. Our findings suggest that peer influence can both reinforce protective behaviors and normalize risky practices, underscoring the critical need for peer-focused designs and educational programs that leverage social connections to guide youth toward safer online practices. By understanding the social dynamics that shape youth’s digital decisions, designers and educators can create tools and resources that empower young users to navigate the digital landscape safely, without sacrificing their need for connection and affirmation.

\begin{acks}
This research is supported in part by the U.S. National Science Foundation under grants \#IIP-2329976, \#IIS-2333207 and by the William T. Grant Foundation grant \#187941. Any opinions, findings, and conclusions or recommendations expressed in this material are those of the authors and do not necessarily reflect the views of the research sponsors. We would also like to thank all the participants who donated their data and contributed towards our research.
\end{acks}

\bibliographystyle{ACM-Reference-Format}
\bibliography{07_References}


\end{document}